# A Hybrid MPI-CUDA Approach for Nonequispaced Discrete Fourier Transformation

Sheng-Chun Yang [a,*], Yong-Lei Wang [b]

[a] School of Computer Science, Northeast Electric Power University, Jilin 132012, China
[b] Department of Materials and Environmental Chemistry, Arrhenius Laboratory, Stockholm University, SE-106 91, Stockholm, Sweden
E-mail: foxysc@foxmail.com; wangyonl@gmail.com

**Abstract:** Nonequispaced discrete Fourier transformation (NDFT) is widely applied in all aspects of computational science and engineering. The computational efficiency and accuracy of NDFT has always been a critical issue in hindering its comprehensive applications both in intensive and in extensive aspects of scientific computing. In our previous work (2018, S.-C. Yang *et al.*, *Appl. Comput. Harmon. Anal.* 44, 273), a CUNFFT method was proposed and it shown outstanding performance in handling NDFT at intermediate scale based on CUDA (Compute Unified Device Architecture) technology. In the current work, we further improved the computational efficiency of the CUNTTF method using an efficient MPI-CUDA hybrid parallelization (HP) scheme of NFFT to achieve a cutting-edge treatment of NDFT at super extended scale. Within this HP-NFFT method, the spatial domain of NDFT is decomposed into several parts according to the accumulative feature of NDFT and the detailed number of CPU and GPU nodes. These decomposed NDFT subcells are independently calculated on different CPU nodes using a MPI process-level parallelization mode, and on different GPU nodes using a CUDA thread-level parallelization mode and CUNFFT algorithm. A massive benchmarking of the HP-NFFT method indicates that this method exhibit a dramatic improvement in computational efficiency for handling NDFT at super extended scale without loss of computational precision. Furthermore, the HP-NFFT method is validated via the calculation of Madelung constant of fluorite crystal structure, and thereafter verified that this method is robust for the calculation of electrostatic interactions between charged ions in molecular dynamics simulation systems.

**Keywords:** NDFT; MPI; GPU; CUDA; CUNFFT; HP-NFFT; Madelung constant; Electrostatic interactions; Molecular dynamics simulations.

## 1. Introduction

It is known that the forward discrete Fourier transform (DFT) and its inverse procedure (IDFT) are described by discrete summation of arrays in spatial and temporal domains, respectively, and are written as:

$$\hat{f}_k = \sum_{j=0}^{N-1} f_j \cdot e^{-\frac{2\pi i k j}{N}} (k = 0, \cdots, N\text{-}1), \tag{1}$$

$$f_j = \frac{1}{N}\sum_{k=0}^{N-1} \hat{f}_k \cdot e^{\frac{2\pi i k j}{N}} (j = 0, \cdots, N\text{-}1), \tag{2}$$

where $N \in \mathbb{N}$, and coefficients $f_j, \hat{f}_k \in \mathbb{C}$. It is obvious that the direct computation of both DFT and IDFT requires $O(N^2)$ arithmetic operation, whereas the fast Fourier transform (FFT) method can effectively scaling down their arithmetic operations to $O(N\log N)$ [1].

A generalization of *d*-dimensional DFT can be described by an array of equispaced spatial points and a certain frequency bandwidth vectors ***N***, in which each component $N_t \in 2\mathbb{N}$ ($t = 0, \cdots, d\text{-}1$) is defined as a fixed even number. These components are collected in the vector ***N***:=$(N_0, \cdots, N_{d-1})^T$, and the corresponding inverse counterparts are determined via ***N***$^{-1}$:= $(1/N_0, \cdots, 1/N_{d-1})^T$, respectively. The set $I_N := \{\boldsymbol{k} \in \mathbb{Z}^d: -N_t/2 \le k_t < N_t/2, t=0, \cdots, d\text{-}1\}$ is an ensemble of all possible *d*-dimensional frequency coordinates in a DFT matrix, where $k_t$ is a component of ***k*** at dimension *t*. In this way, the equispaced points in DFT can be depicted as $\boldsymbol{x}_j = \boldsymbol{N}^{-1} \odot \boldsymbol{j}$ ($\boldsymbol{j} \in I_N$), where the operator $\odot$ does component-wise product of two vectors. Therefore, it is evident that the data points $(j_0/N_0, j_1/N_1, \cdots, j_{d-1}/N_{d-1})$ in DFT are regularly distributed in a spatial domain $\Pi^d$, $\Pi^d := [-0.5, 0.5)^d$ and the number of these data points is $|I_N| = \prod_{t=0}^{d-1} N_t$. Thus, the *d*-dimensional DFT and IDFT shown in Eq. (1) and Eq. (2), respectively, can be rewritten as,

$$\hat{f}(\boldsymbol{k}) = \sum_{\boldsymbol{j} \in I_N} f(\boldsymbol{x_j}) \cdot e^{-2\pi i \boldsymbol{k} \boldsymbol{x_j}}, \tag{3}$$

$$f(x_j) = \frac{1}{|I_N|}\sum_{k \in I_N} \hat{f}(k) \cdot e^{2\pi i k x_j}, \tag{4}$$

where $x_j = N^{-1} \odot j$ and $j, k \in I_N$. With such description, the direct computation for $d$-dimensional DFT (Eq. (3)) and IDFT (Eq. (4)) generally demands for a computational amount of $O(|I_N|^2)$ arithmetic operation, which will be further scale down to $O(|I_N| \log |I_N|)$ using FFT.

The nonequispaced discrete Fourier transform (NDFT) [2] can be viewed as a generalization of DFT with the data points $x_j$ ($j \in [0, M-1]$) randomly scattered in the spatial domain $\Pi^d := [-0.5, 0.5)^d$ and the number of data points $M$ is arbitrary (need not be equal to $|I_N|$). Assuming that the Fourier coefficients $f(x_j) \in \mathbb{C}$ are values at nonequispaced points $x_j$ ($j \in [0, M-1]$) and $\hat{f}(k) \in \mathbb{C}$ ($k \in I_N$) are the output values of this transform, the NDFT is defined as

$$\hat{f}(k) = \sum_{j=0}^{M-1} f(x_j) \cdot e^{-2\pi i k x_j} \quad (k \in I_N). \tag{5}$$

Eq. (5) can be denoted by a matrix-vector product of $\hat{f} = A \cdot f$, where $\hat{f} := (\hat{f}(k))_{k \in I_N}$ is the transform output results, $A := (e^{-2\pi i k x_j})_{k \in I_N, j=0,\cdots,M-1}$ is the nonequispaced Fourier matrix, and $f := (f(x_j))_{j=0,\cdots,M-1}$ is the input data, respectively. Unlike DFT and IDFT, the matrix $A$ in NDFT is typically not square. Even it is square by coincidence, its inverse and orthogonal matrices are usually not existent. Therefore, the inverse NDFT process can be customarily defined by a matrix-vector product of $f = A^H \cdot \hat{f}$ with matrix $A^H := (e^{2\pi i k x_j})_{j=0,\cdots,M-1; k \in I_N}$, and the inverse format of NDFT is described as

$$f(x_j) = \sum_{k \in I_N} \hat{f}(k) \cdot e^{2\pi i k x_j} \quad (j = 0, \cdots, M-1). \tag{6}$$

It is noteworthy that both NDFT and its inverse process are widely applied in all aspects of engineering areas and scientific computation. They play a dominant role in many specific applications when it comes to medical imaging technology [3, 4], astrophysics [5], oceanography [6], remote sensing technology [7, 8], telecommunication [9-11], power system analysis [12-16], and molecular dynamics simulations [17-21].

Due to the arbitrary number of data points and their random distributions in NDFT, it is not feasible to use FFT algorithm to directly handle NDFT (Eq. (5)) and its inverse format (Eq. (6)). In addition, the direct calculation of NDFT and its inverse process demands for approximately $O(M|I_N|)$ arithmetic operations, which is too expensive to accomplish in practical applications. Therefore, several strategies have been proposed to accelerate the computation of NDFT and its inverse process, of which the most prestigious one is nonequispaced fast Fourier transform (NFFT) technique [22]. There are some other names for NFFT, which are, for example, generalized FFT [23], unequally-spaced FFT [24], non-uniform FFT [25], fast approximate Fourier transforms for irregularly spaced data [26], *etc*. Steidl introduced an approximation approach to evaluate computational efficiencies of representative NFFT algorithms in detail [26]. Later, Benedetto and Ferreira proposed a unified approach to perform fast evaluation of NDFT and its inverse process [27].

In our previous work, we proposed a novel and efficient theoretical derivative of NFFT based on gridding algorithm, and implemented in GPU framework using CUDA (Compute Unified Device Architecture) parallel technology and named as CUNFFT [28]. The detailed physical parameters of CUNFFT algorithm including variation in window function, structure of data storage, memory allocation, and thread schedule *etc.* are systematically evaluated to maximize the functional performance of CUNFFT algorithm. In comparison with CPU version of NFFT, the CUNFFT algorithm does not resort to a strategy of precomputation to avoid a time-consuming access to global memory. A series of benchmarking calculations shown that CUNFFT exhibits distinct performance over conversional version at the same computational precision. It should be noted that although CUNFFT gives a favorable efficiency over conventional CPU version of NFFT in representative applications, the limit storage and computational capability of a single GPU node may affect the efficiency of CUNFFT in handling computational systems at super extended scales. In such case, the mass data have to be divided into several groups according to different capabilities of GPU nodes in treating matrix data and input these group data into GPU one after another instead of calculating all matrix data in a whole process. In the current work, we proposed an effective parallelization procedure to further improve computational efficiency of CUNFFT algorithm. Multiple GPU nodes are integrated in a network architecture, and the CUNFFT algorithm is restructured using an efficient hybrid parallelization (HP) scheme unifying MPI (Message Passing Interface) and CUDA technologies [29]. The proposed HP-NFFT method is first validated via the calculation of Madelung constant of fluorite crystal structure, and thereafter is verified to be robust for treatment of matrix data at super extended scale with cutting-edge functional performance and for the calculation of electrostatic interactions between charged ions in molecular dynamics simulation systems.

## 2. The CUNFFT algorithm

The basic principle of CUNFFT algorithm is that an approximation scheme is first contrived based on a specific window function $\varphi$ [30, 31] and thereafter standard FFTs are incorporated into this scheme to evaluate NDFT and its inverse process. For $x \in \Pi^d, k \in I_N$, and $f, \hat{f} \in \mathbb{C}$, the fast and robust evaluation of NDFT (Eq. (5)) and the inverse NDFT (Eq. (6)) are the main objectives. Without loss of generality, the CUNFFT algorithm is described by a 2-dimensional case for the convenience of expression as discussed in the following part.

First, we focus on the fast computation of NDFT (Eq. (5)). As variants $x_j$ in NDFT matrix (Eq. (5)) are irregularly distributed within space $\Pi^d$, the FFT algorithm can't be applied directly to handle NDFT matrix. In order to introduce FFT into NDFT, an approximation scheme of CUNFFT is contrived using the gridding algorithm [32-35] based on Gaussian window function $\varphi$. The adopted window function is alterable, but must be well localized both in spatial and in frequency domains. The combination of FFT and gridding algorithms can be described in three principle steps as illustrated in Fig. 1: first spreading the contribution of $f(x_j)$ on nonequispaced points into the equispaced cells to obtain the value of $g(l)$ at cell $l$ ($l \in I_n$) in spatial domain, thereafter executing Fourier transform in equispaced cells using FFT to obtain the value of $\hat{g}(k)$ at location $k$ ($k \in I_n$) in frequency domain, and finally scaling FFT results to obtain $\hat{f}(k)$ on the appointed points $k \in I_N$.

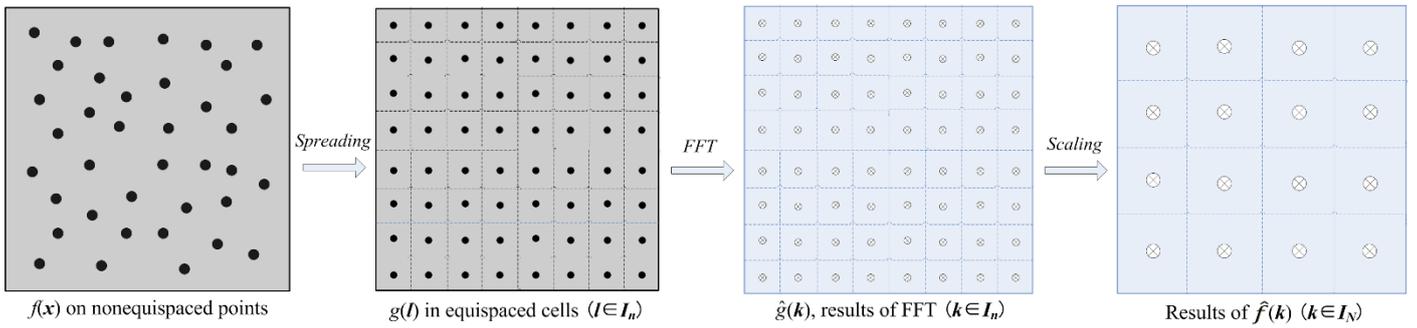

Figure 1. The detailed gridding and FFT procedures for CUNFFT algorithm.

Second, the computation of inverse NDFT (Eq. (6)) is achieved using an approximation scheme and the inverse FFT algorithm. The contribution of values of $\hat{f}(k)$ ($k \in I_N$) is first mapped into $\hat{g}(k)$ in cells at finer scale with $k \in I_n$ for a high computation precision. Thereafter, the values of $\hat{g}(k)$ ($k \in I_n$) in frequency domain are transformed into values of $g(l)$ ($l \in I_n$) in spatial domain using inverse FFT. In the end, the computational results of inverse NDFT are obtained via the interpolation on the given irregular points $x_j$ using Gaussian window function $\varphi$. A schematic procedure for the inverse CUNFFT is illustrated in Fig. 2.

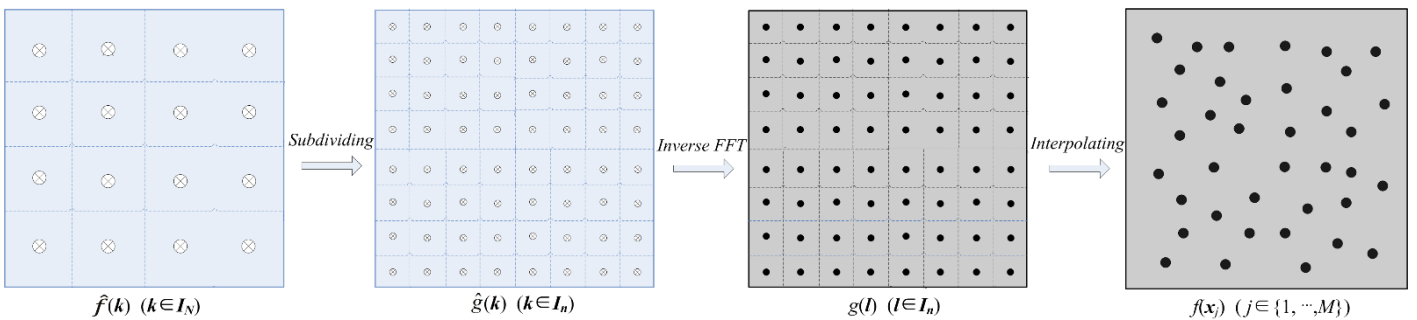

Figure 2. The detailed procedures of gridding algorithm for inverse CUNFFT.

In order to further improve the computational efficiency, the detailed implementation of CUNFFT and inverse CUNFFT algorithms is achieved based on CUDA technology. The CUDA programming model is designed for data-parallel processing to speed up computations, and all CUDA threads are executed on a physically separate *device* (GPU) cooperating with the CPU *host* to run CUDA C and C programs. The parallel code of CUDA threads is executed on GPU nodes and the C program is executed on CPU nodes, respectively. Such a procedure maps data elements to parallel processing *threads* by kernel functions, in which a kernel is adopted to run $N$ times using different data elements in $N$ different CUDA threads with high arithmetic intensity. It is know that in the CUDA programming model both CPU *host* and GPU *device* maintain their own separate memory space in DRAM, referred to *host* memory and *device* memory, respectively. The data transfer between *host* and *device* memory is executed via CUDA runtime functions.

All these steps in CUNFFT algorithm have been highly optimized in GPU nodes with CUDA technology and computations of FFT and inverse FFT in above mentioned procedures taking advantage of internal parallel library functions of CUDA CUFFT on GPUs. The detailed theoretical descriptions and implementation procedures are described in our recent work in Ref. [28].

## 3. The strategy and implementation of HP-NFFT

Even there is a huge boost on the computing power and capability of NVIDIA GPU nodes, the storage capability of a single GPU node is still limited. From the logical architecture, several concepts of *thread*, *block* and *grid* are useful for the understanding CUDA programming model. There are two levels of parallelism in the CUDA programming model, *thread parallelism* and *block parallelism*, which correspond to fine-grained and coarse-grained data parallelisms respectively. The *thread parallelism* is nested within the *block parallelism*. A task corresponding to a *grid* is divided into several sub-tasks that are solved independently in parallel by *blocks* of threads and each sub-task further into finer pieces that are solved cooperatively in parallel by *threads* within a *block*, as illustrated in Fig. 3. Actually, each *block* of threads can be partitioned to any of the available Streaming Multiprocessors (SMs) within a GPU node, either concurrently or sequentially, so that a compiled CUDA program can be executed on a GPU node with any number of SMs. There are multiple Streaming Processors (SPs) within a SM. The number of *threads* in a *block* is typically equal to the number of SPs in an SM. The number of *block*s in a *grid* is usually determined by the size of data matrix to be processed, which is often far more than the number of SMs within a GPU node.

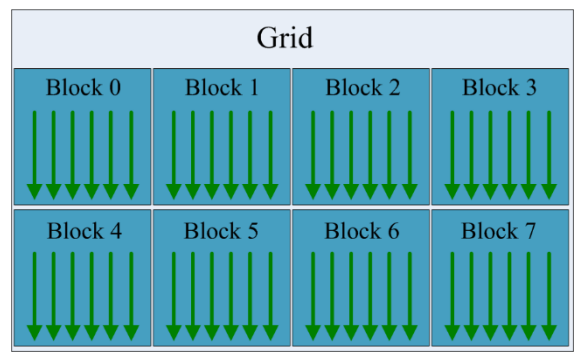

Figure 3. A task corresponds to a grid and is composed of thread blocks. Green arrows represent threads within a block.

The maximum number of parallel threads in a GPU node is determined by the computing power of this specific GPU node. For a NFFT treatment system at super extended scale, the number of threads required by the CUNFFT algorithm is always far more than the number of SPs in a GPU node. Therefore, the number of threads has to be divided into several groups according to the number of SPs in each SM and the computational data within threads will be executed by groups. In this case, the computing pattern between groups is serial, which will severely affect the computational efficiency of CUNFFT algorithm.

From the physical architecture, there are also three levels of memories in a GPU node, *local memory*, *shared memory* and *global memory* as illustrated in Fig. 4. A thread in CUDA can access to computational data from any level of memories during their execution. A thread has its own *local memory* with the highest access efficiency. Each *block* has *shared memory* that is visible to all threads within this *block*. The *shared memory* in each block has lower efficiency than the *local memory*. In addition, all threads in a GPU node can access to the *global memory* with the lowest efficiency. In general, the higher efficient memory has less data storage space and *vice versa*. Therefore, a reasonable allocation of memory with different access efficiencies is a critical procedure to achieve optimized execution efficiencies in handling specific computational systems.

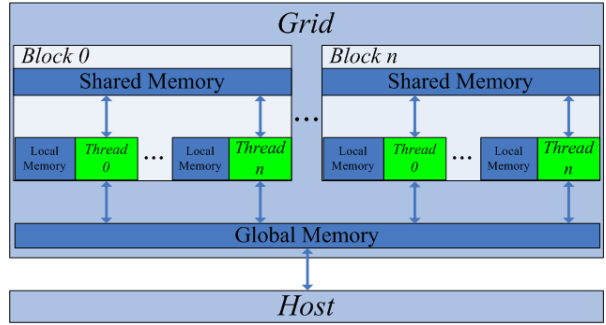

Figure 4. The hierarchical memory structures in a physical architecture. There are three types of memories (local memory, shared memory and global memory) in a GPU node. The CPU *host* communicates with GPU *device* via global memory (DRAM).

For a NFFT treatment system at super extended scale, a single GPU node with limited *local memory* and *shared memory* is not qualified for an efficient computation of NDFT and its inverse process using the CUNFFT algorithm. This indicates that the CUNFFT algorithm has to put large quantity of data into *global memory* in a heave queue and waiting for the calculation and frequent swap of computational data between GPU and CPU nodes, which will significantly affect computational efficiency of CUNFFT algorithm in handling a NFFT treatment system at super extended scale.

Therefore, in order to minimize deficiencies as mentioned in logical and physical architectures, we propose a hybrid parallelization (HP) computational strategy of NFFT based on MPI and CUDA technologies to improve the computational efficiency of the CUNFFT algorithm in handling NDFT at super extended scale. The HP-NFFT method adopts a hybrid parallelization scheme consisting of *CPU parallelism* and *GPU parallellism*. This method is first paralleled via spatial-domain decomposition approach which is implemented by MPI libraries on multiple computer nodes. In this step, the spatial domain of NDFT matrix is decomposed into several subcells with same size. Thereafter, the HP-NFFT method is paralleled via threads within GPU nodes. Each GPU node will concurrently handle all data points in the subcell determined in the first step. A representative physical architecture of this hybrid parallelization scheme is illustrated in Fig. 5.

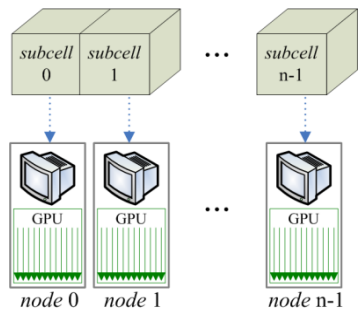

Figure 5. The representative architecture of the hybrid parallelization scheme.

In HP-NFFT method, the computation of NDFT and its inverse process will be executed on multiple GPU nodes distributed in different computer nodes. Thus, the key issue is the divisibility of NDFT, which is critical in logically determining the feasibility of HP-NFFT method. Taking advantage of the linear accumulation feature of NDFT, the transformation results $\hat{f}(\mathbf{k})$ in Eq. (5) can be expressed as a summation of several smaller NDFTs. As illustrated in Fig. 6, the $N$ nonequispaced points in a computational system are splitted into $m$ (herein $m = 4$) subcells. There are approximately $C = N/m$ points to be handled in each subcell, thus the computation of $\hat{f}(\mathbf{k})$ can be decomposed into $m$ accumulative terms as described in Eq. (7).

$$\hat{f}(\mathbf{k}) = \sum_{i=0}^{C-1} \exp(-\frac{2\pi \mathbf{i}}{L} \mathbf{k} \cdot \mathbf{x}_i) + \sum_{i=C}^{2C-1} \exp\left(-\frac{2\pi \mathbf{i}}{L} \mathbf{k} \cdot \mathbf{x}_i\right) + \cdots + \sum_{i=(m-1)C}^{N-1} \exp(-\frac{2\pi \mathbf{i}}{L} \mathbf{k} \cdot \mathbf{x}_i) \qquad (7)$$

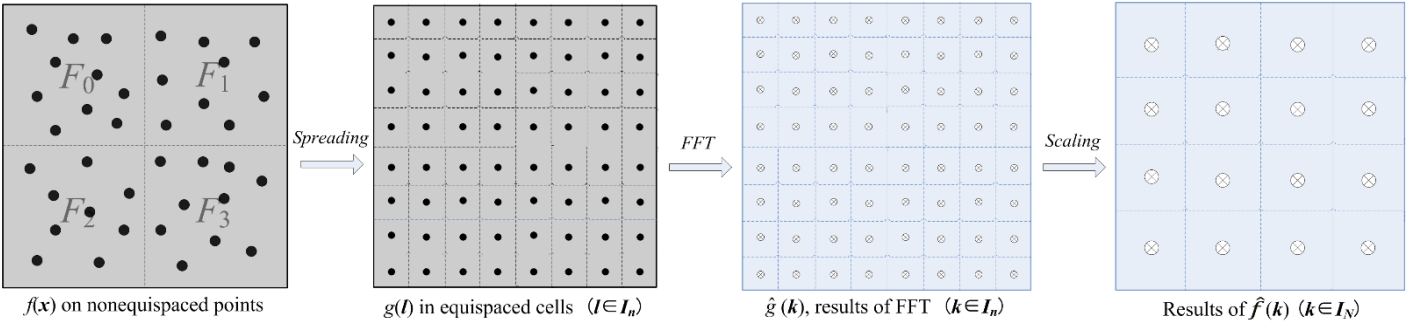

Figure 6. A schematic procedure of gridding algorithm for CUNFFT in a 2-dimensional case. In this decomposition procedure, the NDFT matrix is decomposed into 4 subcells that will be to be handled individually using different GPU nodes.

For simplicity, Eq. (7) can be expressed as Eq. (8), in which each $\hat{f}_i(\bm{k})$ corresponds to a sub NDFT matrix for a specific subcell $i$ that will be calculated on a GPU node. The computation of $\hat{f}_i(\bm{k})$ for each sub NDFT matrix is analogous to CUNFFT but only with less input information of data points in subcell $i$, as illustrated in Fig. 7. After computation of $\hat{f}_i(\bm{k})$ on different GPU nodes, all output data of $\hat{f}_i(\bm{k})$ are collected in a given node (for example computer node 0), in which $\hat{f}(\bm{k})$ is obtained by a summation of all $\hat{f}_i(\bm{k})$ subsets as shown in Eq. (8). Owing to the small size of $\hat{f}_i(\bm{k})$, the cost of transferring data between computer nodes is acceptable in comparison with the overall improvement of computational efficiency of NDFT computation. Therefore, the computational scheme of HP-NFFT method consists of one decomposition procedure and several CUNFFTs. These CUNFFTs are conducted concurrently in handling all data points in the corresponding subcells in different GPU nodes. This is the so-called *GPU parallelism*.

$$\hat{f}(\bm{k}) = \hat{f}_0(\bm{k}) + \hat{f}_1(\bm{k}) + \cdots + \hat{f}_{m-1}(\bm{k}) \tag{8}$$

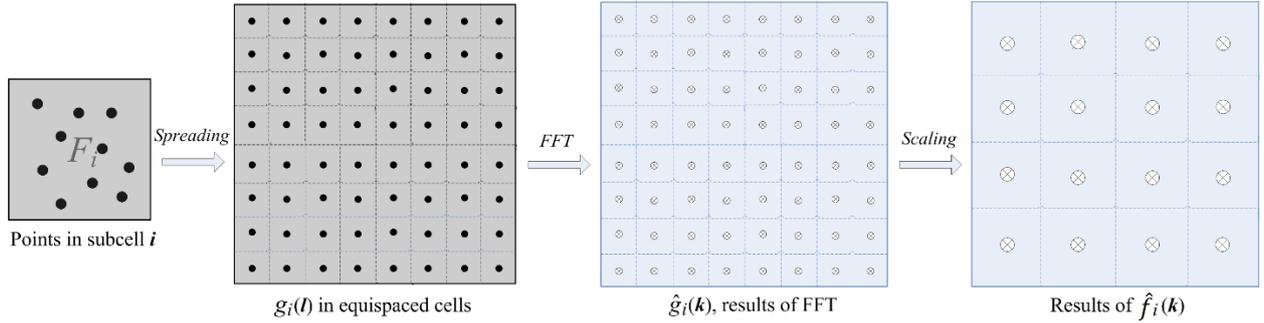

Figure 7. A component computation of subcell in HP-NFFT, which is executed on a single computer node.

After each CUNFFT calculation, all computer nodes will communicate with others and share their respective computational results for the next round CUNFFT calculation. In HP-NFFT method, both the spatial domain decomposition approach and the data collection after each CUNFFT calculation are achieved via the *CPU parallelism*. The *CPU parallelism* procedure in HP-NFFT method is composed of a set of MPI processes, each of which runs on a CPU core responsible for the spatial domain decomposition of NDFT matrix and the data collection after each CUNFFT calculation. This *CPU parallelism* procedure is implemented by MPI libraries where there is a consistent one-to-one matching relationship between a MPI process and a GPU node. The detailed *CPU parallelism* scheme is described in Algorithm 1, in which the computational system (NDFT matrix) is first decomposed into *num_nodes* subcells on the major computer node (*rank* = 0), from which the detailed information of each subcell are subsequentially sent to the corresponding nodes. Each computer node computes the NFFT of its own subcell $\hat{f}_i(\bm{k})$ individually and concurrently via *CompCUNFFT* function, and thereafter the *Accumulate* function will finalize the summation of all output data of $\hat{f}_i(\bm{k})$ ($i = 0, \cdots, m$-1) according to the frequency coordinate $\bm{k}$.

**Algorithm 1:** The main architecture of HP-NFFT method (CPU parallelism)

**Require:** The computational system *information (NDFT matrix)*.
**Require:** The number of computer nodes *num_nodes*.
1. MPI_Init(&argc, &argv);
2. MPI_Comm_rank(MPI_COMM_WORLD, &rank);
3. MPI_Comm_size(MPI_COMM_WORLD, &num_nodes);

```
 4.    if rank == 0 then
 5.        subcells = Split(box);
 6.        subcell = subcells[0];
 7.        for i = 1 to num_nodes-1 do
 8.            MPI_Send(subcells[i], node[i]);
 9.        end for
10.    else
11.        MPI_Recv(subcell, node[0]);
12.    end if
13.    $\hat{f}_{rank}(\boldsymbol{k})$ = CompCUNFFT(subcell);
14.    Accumulate($\hat{f}_{rank}(\boldsymbol{k})$);
15.    MPI_Finalize();
```

It is noted that the CUNFFT algorithm is intrinsically a thread-level parallel CUDA function, by which the *GPU parallelism* is implemented in local GPU node. The process of a CUDA function is composed of three steps: allocating GPU *memories* and loading data from CPU *host* to GPU *device*, executing *kernel* functions for specific tasks, and retrieving output data from GPU *device* to CPU *host*. The detailed implementation procedure of CUNFFT is shown in Algorithm 2. The nonequispaced coordinates $x[j]$ and their values $f[j]$ in subcell $i$ are transferred from *host memory* to the allocated *global memory* in GPU *device*. Thereafter, the three steps of CUNFFT algorithm, *Spreading*, *FFT* and *Scaling*, are conducted one after another. Both *Spreading* and *Scaling* steps are implemented by CUDA parallel mechanism in *spread* and *scale* kernel functions, respectively, and can be performed within GPU *device* in parallel. The *FFT* step is conducted using the CUFFT function, which is an internal kernel function of CUDA. In the last step, the transformation output data $\hat{f}_i(\boldsymbol{k})$ of subcell $i$ are sent back to CPU *host* and the allocated *global memories* will be released.

**Algorithm 2:** The CUNFFT for fast computation of NDFT (GPU parallelism)

**Require:** Frequency domain $I_N$

**Require:** $x_j \in$ subcell $i$, $f(x_j) \in \mathbb{C}$

1. **Load data from host to device**
2. **Execute** *Spreading kernel*
   spread<<<block_num, thread_num>>>(params)
3. **Execute** *Forward CUFFT*
4. **Execute** *Scaling kernel*
   scale<<<block_num, thread_num>>>(params)
5. **Retrieve results from device to host**

**Output:** approximate values $\hat{f}(k)$, $k \in I_N$

In the *Spreading* step of CUNFFT, the value of $g_i(\boldsymbol{l})$ for each regular lattice $\boldsymbol{l}$ is obtained by a summation of the contributions of points within the neighborhood $J(\boldsymbol{l})$ of lattice $\boldsymbol{l}$. The contribution of a point to a neighbor lattice is related to the distance between them. An increase in the relative distance between the point and the lattice leads to a smaller contribution of the point to the lattice, and *vice versa*. If the relative distance between the point and the lattice is larger than a threshold value, the contribution of the point to the lattice will be ignored. Therefore, a lattice can restrict its neighborhood $J(\boldsymbol{l})$ to a small space for effective computation. Due to the irregular distribution of data points in main NDFT matrix and NDFT subsets, the number of data points within neighborhood of each lattice differs dramatically, which would result in a severe unbalance of parallel computation. Therefore, the computational focus should be converted from lattice $\boldsymbol{l}$ to nonequispaced point $x_j$, which is taken as a central point to calculate the contributions to all lattices within its neighborhood $I(x_j)$. Such a treatment enables the amount of calculation for *Spreading* step being balanced on $C$ threads. Each thread is responsible for one point $x_j$ ($j$=0, ···, $C$-1) to account for its contribution to neighboring lattices, and therefore the performance of parallel computation of CUNFFT in GPU node is significantly enhanced. It should be noted that all threads in GPU nodes concurrently conduct parallel transformations and then determine the contributions of all points to neighboring lattices, which may result in mutex error when multiple threads did similar operation on the same lattice at the same time. In order to avoid potential mutex error, we introduce an atomic operation function *AtomicAdd()* to consequently accumulate output data from all other threads to $g_i(\boldsymbol{l})$.

For kernel functions in Algorithm 2, the parameter *thread_num* corresponds to the number of threads in each *block*. Then, the number of *block*s (*block_num*) in a *grid* can be obtained via the total number of threads in the *grid* divided by the number of threads in each block (*thread_num*). For example there are 1024 nonequispaced points in a subcell, meaning that there should be 1024 threads in a *grid* to handle the contributions of all these nonequispaced points to neighboring lattices. If one *block* is defined to accommodate 256 threads (*thread_num* = 256) at most, all threads (1024) in a *grid* will be divided into 4 *block*s (*block_num* = 4). All *block*s within a *grid* and all threads within a *block* can be identified by the built-in variables *blockIdx* and *threadIdx*, respectively. More specifically, the *block*s in the *grid* mentioned above can be identified by *blockIdx.x* ($x$ = 0, $\cdots$, 3) and the *threads* in a *block* can be indentified by *threadIdx.x* ($x$ = 0, $\cdots$, 255), respectively. Thus, in the *Spreading* step, a thread corresponds to a point $x_j$ and can be identified in GPU memory by a unique index derived from *blockIdx* and *threadIdx*. Each thread conducts the same operation to estimate the contribution of a point $x_j$ to all neighboring lattices if these lattices are located within a threshold value of the central point $x_j$. The *block*s are automatically loaded onto all *SM*s within a GPU node for parallel computation. However, it should be noted that the number of *SM*s varies in different GPU nodes. The more *SM*s within a GPU node, the more *block*s are available to perform parallel computations, which is an important factor to improve the computational efficiency of different GPU nodes as illustrated in Fig. 8.

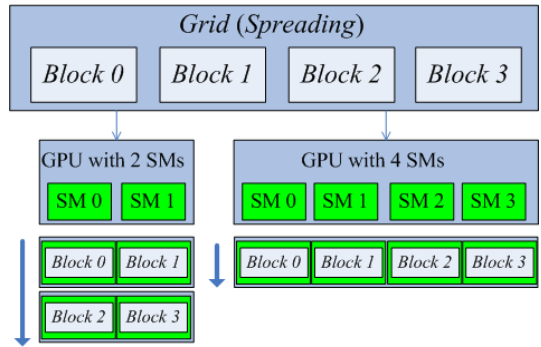

Figure 8. Execution of CUNFFT parallel computation using GPU nodes with different number of streaming multiprocessors. Herein, the *Spreading* kernel is decomposed into 4 blocks of threads that execute independent CUNFFT. A GPU node with 4 streaming processors will take less time than that for a GPU node with 2 streaming processors in performing CUNFFT parallel computations.

In the *FFT* step of CUNFFT, the values of $g_i(l)$ on equispaced spatial lattices $l$ ($l \in I_n$) are transformed into $\hat{g}_i(k)$ in the equispaced frequency domain $k$ ($k \in I_n$) using FFT algorithm. This transformation process is conducted in GPU nodes in parallel using a built-in kernel function *CUFFT*. In addition, this built-in kernel function can effectively avoid computational times for transferring data between CPU *host* and GPU *device*.

In the *Scaling* step of CUNFFT, the equispaced frequency domain of $\hat{f}_i(k)$ ($k \in I_N$) generally differs from that of $\hat{g}_i(k)$ ($k \in I_n$) due to the adoption of an oversampling factor in the *Spreading* step for consistent computational accuracy. In fact, the frequency domain $I_N$ for $\hat{f}_i(k)$ is smaller than the frequency domain $I_n$ for $\hat{g}_i(k)$, that is, $I_N \subset I_n$. Therefore, the values of $\hat{f}_i(k)$ can be obtained from $\hat{g}_i(k)$ divided by a weight factor $c_k(\tilde{\varphi})$. The weight factor $c_k(\tilde{\varphi})$ is obtained from a periodic window function $\tilde{\varphi}$ in frequency domain $k$ ($k \in I_N$). The parallel execution strategy of the *Scaling* kernel is similar to that in the *Spreading* kernel. Each thread corresponds to one equispaced lattice in frequency domain indexed by value $k$ ($k \in I_N$), and is responsible for the calculation of $\hat{f}_i(k)$.

The *Accumulate* function in Algorithm 1 is used to perform an overall summation of output data for all subcells of NDFT matrix shown in Eq. (8). These computational results $\hat{f}_i(k)$ ($i$ = 0, $\cdots$, *num_nodes*-1) for all subcells are obtained from different computer nodes and are transferred to the major computer node with *rank* = 0. In addition, a tree-like addition structure is designed, as shown in Fig. 9. This strategy of summation can effectively reduce the computational time from $n$ to $\log n$ for transferring data from all other computer nodes to the major computer node with *rank* = 0. This tree-like addition structure is essentially a CPU parallelism, and therefore the *Accumulate* function is implemented using MPI parallel technology and the detailed procedures are shown in Algorithm 3.

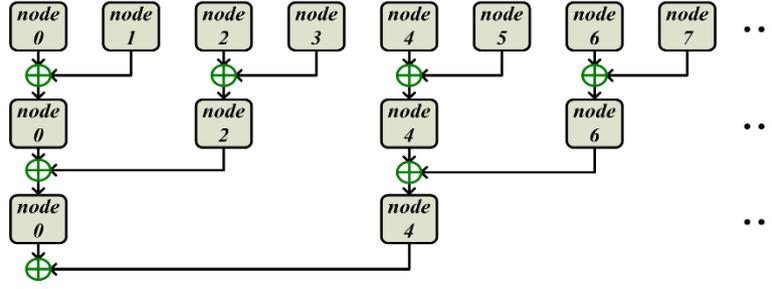

Figure 9. A tree-like addition structure is designed for overall summations in the *Accumulate* function.

---

**Algorithm 3**: The code of *Accumulate* (CPU parallelism)

---

**Require:** The component of sub-NDFT $\hat{f}_{\text{rank}}(\boldsymbol{k})$.
**Require:** The number of computer nodes *num_nodes*.

```
1.   int offset = 1, mask = 1;
2.   while(offset < num_nodes) {
3.       if((rank & mask) != 0) {
4.           MPI_Send(f̂_rank(k), node[rank - offset]);
5.       }else{
6.           MPI_Recv(f̂_rank+offset(k), node[rank + offset]);
7.           f̂_rank(k) += f̂_rank+offset(k);
8.       }
9.       offset += offset;
10.      mask = offset + mask;
11.  }
```

The inverse CUNFFT procedure comprises of three steps as mentioned above: *Subdividing*, *Inverse FFT*, and *Interpolating*. All these steps are also implemented by a two-level hybrid parallel strategy composed of *CPU parallelism* and *GPU parallelism*. In the *CPU parallelism* for inverse CUNFFT, each computer node executes the same operation with the same set of data in the first two steps. However in the *Interpolating* step of inverse CUNFFT, the adopted CPU parallelization scheme is distinct to that used in CUNFFT. As illustrated in Fig. 10, the nonequispaced points in NDFT matrix are divided into $m$ ($m = 4$) subcells in the *Interpolating* step, and each computer node is responsible for the computation of data points in one subcell. The detailed code architecture of such a CPU parallelization scheme is shown in Algorithm 4. The major computer node (*rank = 0*) is responsible for dividing the NDFT matrix consisting of nonequispaced points into $m$ subcells and thereafter distributing them to $m$ computer nodes. Each computer node performs the inverse CUNFFT independently on all data points in the subcell using the same initial input $\hat{f}(\boldsymbol{k})$s. After the computation, all output values of $f(x_j)$ ($j \in \{1,\cdots,M\}$) will be sent back to major computer node (*rank* = 0).

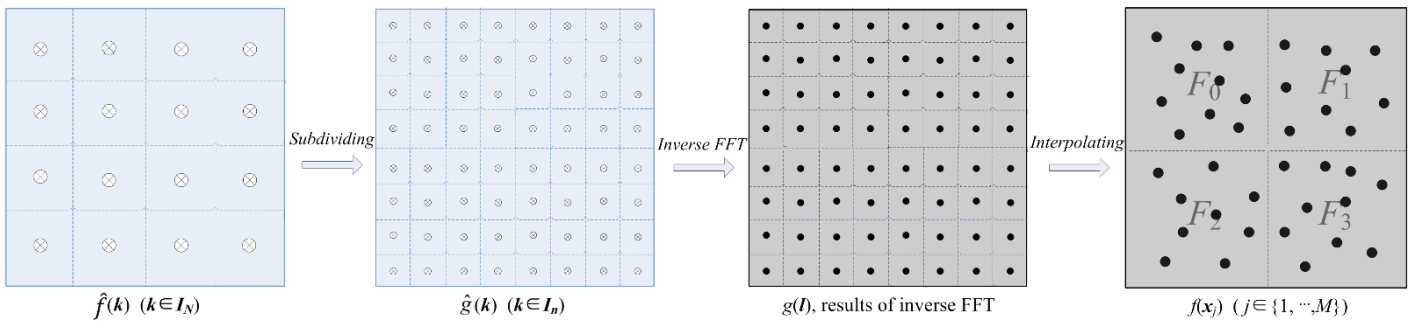

Figure 10. The gridding algorithm used for inverse CUNFFT in a 2-dimensional case. All nonequispaced points in the spatial domain of a NDFT matrix are divided into $m=4$ subcells and the inverse transformation of all data points in each subcell will be calculated in different computer nodes.

---

**Algorithm 4:** The main architecture of inverse CUNFFT in HP-NFFT (CPU parallelism)

---

**Require:** The values of $\hat{f}(\boldsymbol{k})$ in equispaced frequency domain $\boldsymbol{k} \in I_N$.
**Require:** The nonequispaced points $x_j$ ($j \in \{1, \dots, M\}$).

**Require:** The number of computer nodes *num_nodes*.
1. MPI_Init(&argc, &argv);
2. MPI_Comm_rank(MPI_COMM_WORLD, &rank);
3. MPI_Comm_size(MPI_COMM_WORLD, &num_nodes);
4. **if** rank == 0 **then**
5.     subcells = Divide($x_j$); ($j \in \{1, \dots, M\}$)
6.     subcell = subcells[0];
7.     **for** i = 1 **to** num_nodes-1 **do**
8.         MPI_Send($\hat{f}(k)$ & subcells[i], node[i]);
9.     **end for**
10. **else**
11.     MPI_Recv($\hat{f}(k)$ & subcell, node[0]);
12. **end if**
13. $f_{\text{rank}}(x_j \in \text{subcells[rank]})$ = CompInverseCUNFFT($\hat{f}(k)$);
14. **if** rank != 0 **then**
15.     MPI_Send($f_{\text{rank}}(x_j \in \text{subcells[rank]})$, node[0]);
16. **else**
17.     **for** i = 1 **to** num_nodes-1 **do**
18.         MPI_Recv($f_i(x_j \in \text{subcells[i]})$, node[i]);
19.     **end for**
20. **end if**
21. MPI_Finalize();

The *CompInverseCUNFFT* function in Algorithm 4 is used to compute the inverse CUNFFT of all data points in the subcell as illustrated in Fig. 11. In the step of *Subdividing*, the $\hat{f}(k)$s in frequency domain $I_N$ are mapped to $\hat{g}(k)$s in frequency domain $I_n$ ($I_N \subset I_n$). In the step of *Inverse FFT*, the $\hat{g}(k)$s in frequency domain are directly transformed to $g(l)$s in spatial domain using inverse FFT procedures. In the first two steps, each computer node conducts the same operations on the same data. However, in the third step of *Interpolating*, the $f(x)$s are obtained via interpolation of data points if these points belongs to a given subcell. The three steps are implemented by GPU parallelism and the detailed code structure is illustrated in Algorithm 5. First, the *device memory* is allocated to GPU nodes and the input data $\hat{f}(k)$ ($k \in I_N$) are loaded from CPU *host* to GPU *device*. Second, the *Subdividing* kernel performs parallel computations with one thread corresponding to a lattice $k$ and is responsible for the computation of $\hat{g}(k)$ ($k \in I_N$). All values of $g(l)$ are obtained using a GPU built-in kernel function of inverse CUFFT executing on $\hat{g}(k)$ ($k \in I_N$). In the *Interpolating* kernel, each thread corresponds to a data point $x_j$, and is responsible for the computation of $f(x_j)$, which accumulates the contributions of neighbor lattices around this data point $x_j$ to the corresponding $f(x_j)$. It is noteworthy that unlike the *Spreading* kernel, the mutex error in such an accumulation procedure is not available and therefore the atomic operation function is not necessary for the *Interpolating* kernel. Therefore, the inverse HP-NFFT algorithm theoretically has a better performance than the forward HP-NFFT algorithm. After the computation, all output data of $f(x_j)$ ($x_j \in \text{subcells}[rank]$) are sent back from GPU *device* to CPU *host* and the allocated *device memory* will be released immediately.

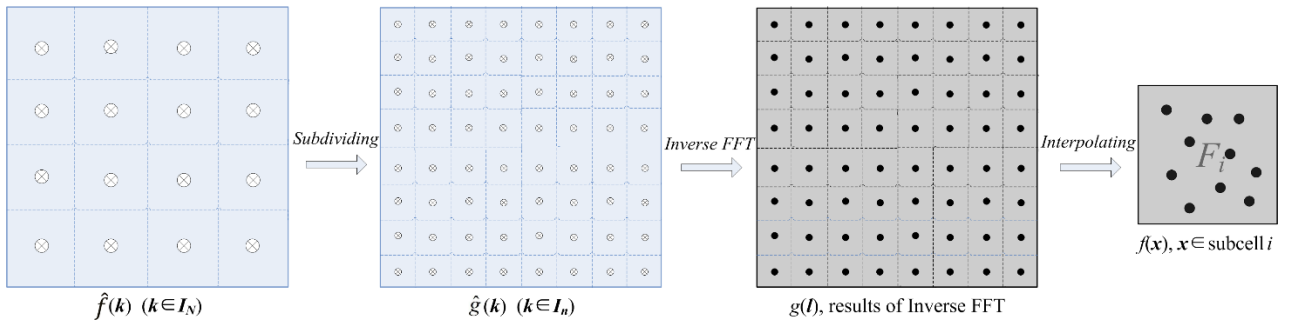

Figure 11. The computation component of a subcell in inverse HP-NFFT algorithm, which is executed on a single computer node.

**Algorithm 5:** Inverse CUNFFT for computation of inverse NDFT (GPU parallelism)

**Require:** The values of $\hat{f}(k)$ in equispaced frequency domain $k \in I_N$.
**Require:** The nonequispaced points $x_j \in \text{subcells}[rank]$.

1. **Load data from host to device**
2. **Execute** *Subdividing kernel*

    *subdivide<<<block_num, thread_num>>>(params)*
3. **Execute** *Inverse CUFFT*
4. **Execute** *Interpolating kernel*

    *interpolate<<<block_num, thread_num>>>(params)*
5. **Retrieve results from device to host**

**Output:** approximate values $f(x_j \in \text{subcells[rank]})$.

## 4. The precision and performance of HP-NFFT

Herein, we take 4096 nonequispaced points $x_j$ and their function values $f(x_j)$ ($j \in [0, 4096)$) in a 3-dimensional space for example. Four computer nodes are adopted and the obtained precision data of HP-NFFT and its inverse process are determined using Eq. (9). The values of $\hat{s}_k$ and $\hat{f}_k$ are computed using direct NDFT and HP-NFFT, respectively. The other physical parameters, *i.e.*, bandwidth components $N_0$, $N_1$, $N_2$=16, oversampling factor $\sigma$ = 2, cut-off factor *m* varying from 1 to 15, are used in the current work, which is similar to that used in our previous works for benchmarking computations [28].

$$E = \frac{\|f-s\|_2}{\|s\|_2} = \left(\sum_{j=0}^{M-1}\left|\hat{f}_k - \hat{s}_k\right|^2 / \sum_{j=0}^{M-1}|\hat{s}_k|^2\right)^{\frac{1}{2}} \tag{9}$$

Four window functions, GAUSSIAN, B_SPLINE, SINC_POWER, and KAISER_BESSEL, are used in the current work, and the corresponding computational precision results with respect to cut-off factor *m* are shown in Fig. 12. These computational results indicate that HP-NFFT and its inverse process exhibit the same precision level as that of CUNFFT [28], indicating that the hybrid parallelization scheme of HP-NFFT doesn't affect the computational precision in handling NDFT matrix.

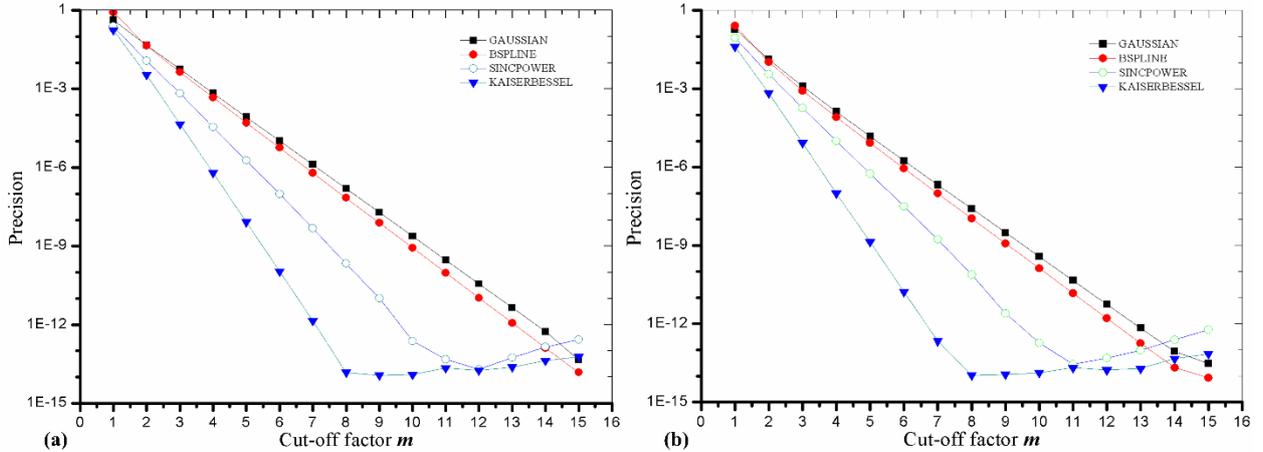

Figure 12. The computational precision of HP-NFFT (a) and its inverse procedure (b) at varied cut-off factor *m* and window functions.

In addition, the performances of CUNFFT and HP-NFFT using varied number of computer nodes are explored. In the current part, as well as additional applications of HP-NFFT method for the calculation of Madelung constant of fluorite crystal structure and the calculation of electrostatic interactions between charged ions in molecular dynamics simulation systems, all computations are performed on the same model computers with configuration of Intel(R) Core(TM) i7-3770 CPU @ 3.40GHz and GTX 980 Ti GPU. The spatial domain of NDFT matrices are composed of $58^3$, $80^3$, $100^3$ and $120^3$ nonequispaced points, respectively.

It is shown in Fig. 13 that the performance of the HP-NFFT method with less computer nodes prevails when the number of points less than $80^3$, otherwise this method with more computer nodes exhibit better performance. This observation indicates that the network delay between different computer nodes has a significant effect on the performance of the HP-NFFT method at given number of nonequispaced points. The high ratio of network delay leads to a low performance of the HP-NFFT method with more nodes when nonequispaced point number less than 80^3. Conversely, when the number of nonequispaced points is larger than a threshold value, the ratio of network delay gets lower and the net execution time dominates the performance of the HP-NFFT method. Therefore, the HP-NFFT method is an efficient parallel algorithm and is suitable for NDFT at super extended scale.

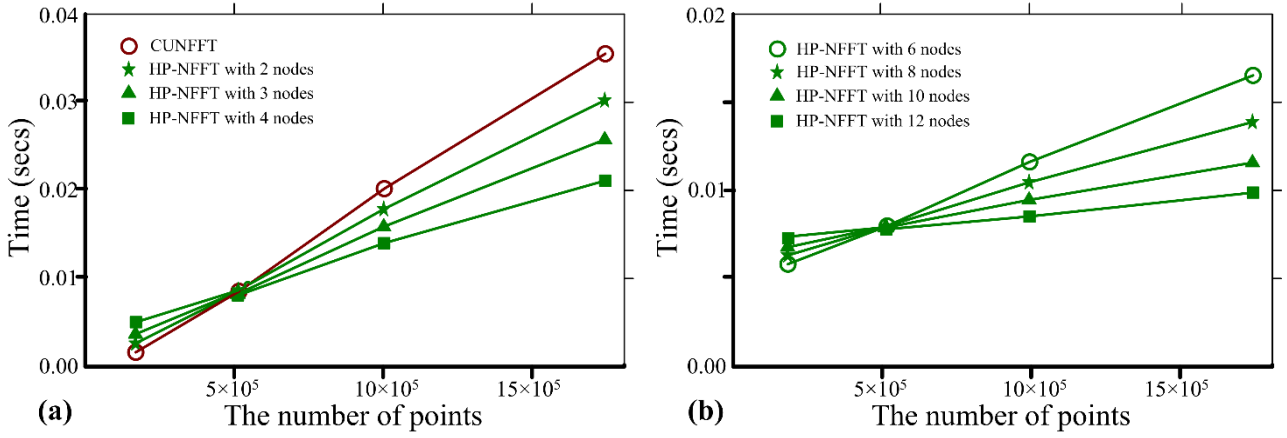

Figure 13.The performance of CUNFFT and HP-NFFT methods in handling NDFT matrices with varied number of nonequispaced points. (a) The performance of CUNFFT using one node and that of HP-NFFT method using computer nodes from 2 to 4. (b) The performance of HP-NFFT method using parallel nodes from 6 to 12.

## 5. Application of HP-NFFT in charged ion systems

In molecular dynamics simulations, the computation of long-range electrostatic interactions between charged particles are the most time-consuming part. The development of efficient methods to calculate electrostatic interactions have always been a challenging task. In past decades, several representative computational strategies [36-40] had been proposed to improve the computation efficiency in treating electrostatic interactions. The Ewald summation method [41] did a remarkable performance in splitting long-range electrostatic interactions into real space and reciprocal space contributions, which laid the foundation for the subsequent particle-mesh Ewald and particle-particle particle-mesh summation methods and their derivatives. In 2006, Laaksonen and coworkers proposed an ENUF (Ewald summation method based on Non-Uniform fast Fourier transform) method for fast calculation of electrostatic interactions between charged ions in molecular simulations systems [17]. In a subsequent study, the ENUF method is further updated using GPU with CUDA technology and combined it with CUNFFT algorithm to improve the computational performance of CU-ENUF method [42].

In the previous sections of this work, we have presented the detailed theory, implementation, and validation parts of the HP-ENUF method. In this section, we use this method to compute the Madelung constant of a specific fluorite crystal to further validate the accuracy of HP-ENUF method in handling long-range electrostatic interactions in molecular systems. The Madelung constant is used in determining the electrostatic potential of a single ion in a crystal structure by approximating ions with point charges [43]. The Madelung constant can be obtained by Eq. (10), where $U^E$ is the total electrostatic energy of a simulation system, $N$ is the total number of ions in the simulation system, $m$ is the number of ions in one molecule, and $Z_+$ and $Z_-$ are the charge valancies of cations and anions, respectively.

$$Madelung = U^E \frac{m}{NZ_+Z_-} \qquad (10)$$

For a crystal structure, its Madelung constant is straightforward once the electrostatic energy $U^E$ is known. According to the ENUF method [17, 18], the electrostatic energy $U^E$ can be decomposed into real-space energy $U^{E,R}$ and reciprocal-space energy $U^{E,K}$, respectively, which are shown in Eq. (11) and Eq. (12).

$$U^{E,R} = \frac{1}{2}\sum_{\mathbf{n}}^{\dagger}\sum_{i,j}\frac{q_iq_j}{|\mathbf{r}_{ij}+\mathbf{n}L|}\operatorname{erfc}(\alpha|\mathbf{r}_{ij}+\mathbf{n}L|), \qquad (11)$$

$$U^{E,K} = \frac{1}{2\pi L}\sum_{\mathbf{n}\neq 0}\frac{e^{-\pi^2|\mathbf{n}|^2/(\alpha L)^2}}{|\mathbf{n}|^2}S(\mathbf{n})S(-\mathbf{n}) - \frac{\alpha}{\sqrt{\pi}}\sum_i q_i^2, \qquad (12)$$

The parameter $L$ represents the side length of the simulation system. The Ewald convergence parameter $\alpha$ in these two equations is used to tune the relative weight between contributions from real space and reciprocal space, but the value of $\alpha$ does not affect the total electrostatic energy [44, 45]. In Eq. (11), the complementary error function is expressed as $\operatorname{erfc}(r) :=$

$\frac{2}{\sqrt{\pi}} \int_r^\infty \exp(-t^2)\, dt$ [46] and the † symbol indicates that in the case of $i=j$, the term $\mathbf{n}=0$ must be omitted. In Eq. (12), $S(\mathbf{n})$ is the structure factor and is denoted by $S(\mathbf{n}) = \sum_i q_i \exp(-\frac{2\pi i}{L}\mathbf{n}\cdot\mathbf{r}_i)$, which is the most time-consuming part in the computation of electrostatic energy. This term is effectively handled in the HP-NFFT method using the MPI-CUDA hybrid parallelization scheme as we proposed in this work.

In a single crystal cell of fluorite ($CaF_2$), as shown in Fig. 14(a), there are 4 $Ca^{2+}$ cations and 8 $F^-$ anions as illustrated by gray and reseda globules, respectively, in Fig. 14(b) [47]. In simulations, we construct a cubic simulation system consisting of $32^3$ fluorite crystal cells and 393216 ions in total. If the side length of a crystal cell is $l$, the shortest distance between $Ca^{2+}$ and $F^-$ is $r_0 = \sqrt{3}l/4$. This value is set as the unit length of simulation system. Therefore, the side length $L$ of the fluorite crystal simulation system can be converted into a reduced dimensionless parameter of $L = 73.9$ used in Eq. (11) and (12). The parameters $q_i$ and $q_j$ in Eq. (11) and (12) is +2 or -1 corresponding to the partial charges of $Ca^{2+}$ and $F^-$ ions, respectively.

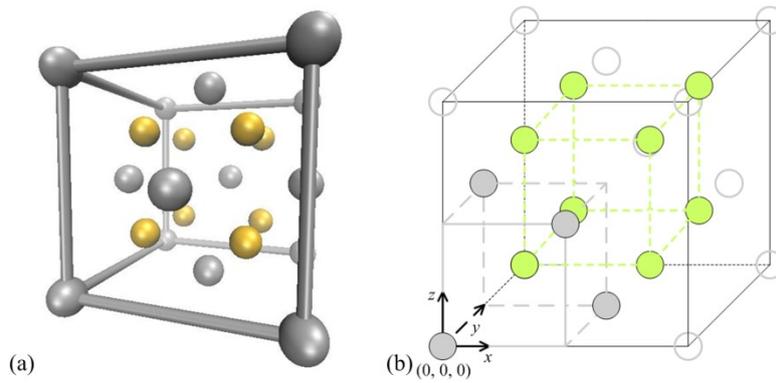

Figure 14. (a) Crystal structure of fluorite; (b) A crystal cell of fluorite, in which gray and reseda balls represent $Ca^{2+}$ and $F^-$, respectively.

The Madelung constant of a fluorite crystal structure is calculated using the HP-ENUF method (ENUF method based on the HP-NFFT method) performed on two computer nodes. It is shown in Fig. 15 that the calculated of Madelung constant values of a fluorite crystal structure converge to theoretical value of 2.5194 [47] for the Ewald convergence parameter α ranging from 1.2 to 1.8. Therefore, the ENUF method in combination with the HP-NFFT method can effectively reproduce the accurate electrostatic energy of a fluorite crystal structure.

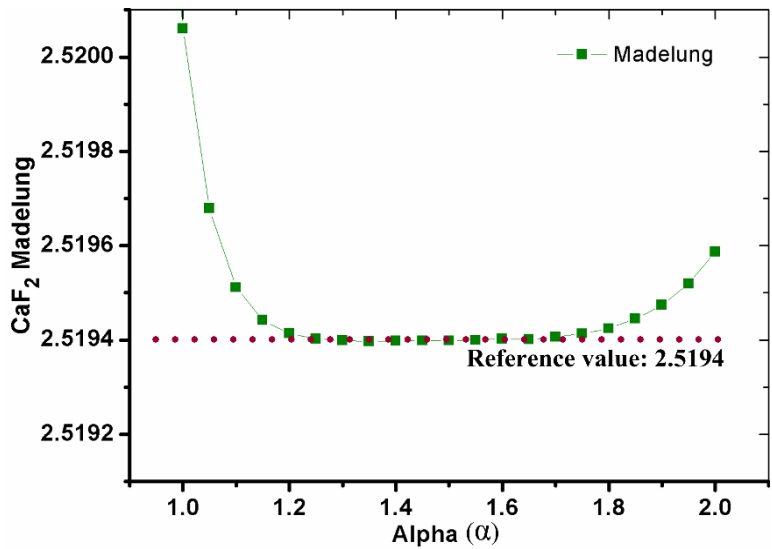

Figure 15. The Madelung constants of a fluorite crystal structure calculated using ENUF and HP-NFFT methods at varied values of Ewald convergence parameter α.

In addition, the calculation of fluorite Madelung constant of fluorite crystal structures are further performed to explore the performance of HP-ENUF method. The fluorite crystal structures are constructed with $12*32^3$, $12*45^3$, $12*56^3$ and $12*64^3$ charged ions, respectively. As shown in Fig. 16, the HP-ENUF method exhibits an obvious computational efficiency in handling long-range electrostatic interactions between charged ions at super extended scale, and present remarkable performance when more computer nodes are used for parallel computation.

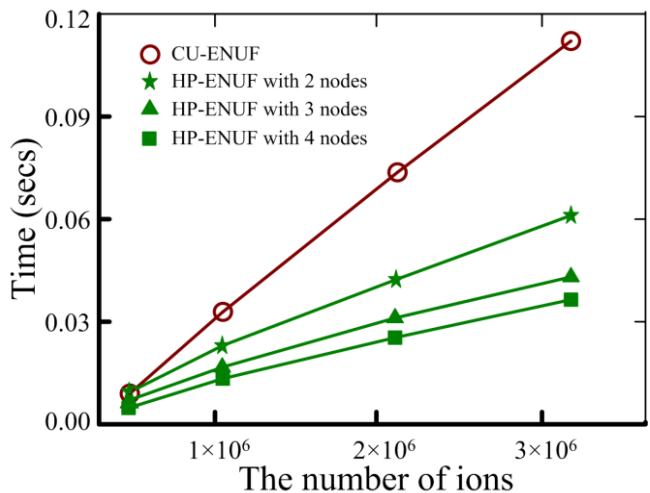

Figure 16. The computational efficiency of HP-ENUF method in calculating electrostatic interactions between charged ions in fluorite crystal structures with varied number of ions. Each simulation system is performed using HP-ENUF method with different computer nodes (CU-ENUF corresponds to the HP-ENUF with one computer node).

These computational results shown in Figs. 15 and 16 indicate that the HP-ENUF method not only has a high precision in the treatment of long-range electrostatic interactions between charged ions with suitable physical parameters, but also exhibits an obvious advantage over CU-ENUF in computational efficiency for simulation systems containing a large number of ions. When a simulation system contains less than $10^6$ charged ions, the HP-ENUF doesn't exhibit its computational efficiency in comparison with CU-ENUF due to distinct network delay among different computer nodes. However, with an increase in the number of charged ions in simulation systems, the HP-ENUF method with more computer nodes shows more superiority of computational efficiency than others. Therefore, the HP-ENUF method is more suitable for the computation of long-range electrostatic interactions between charged ions at super large simulation systems.

**6. Conclusion**
In the current work, we proposed and implemented an HP-NFFT method via a hybrid parallelization scheme using process-level parallelism between CPU nodes and thread-level parallelism in GPU device. As an extension of CUNFFT, the HP-NFFT method exhibits distinct computational efficiency in computing NDFT matrix at a super extended scale. In the HP-NFFT method, the super-scale NDFT matrix is decomposed into several subcells according to the accumulative feature of spatial domain of NDFT matrix. All data points in NDFT matrix are distributed into different computer nodes for the process-level parallel computation, and the computation in each computer node fully inherits the merits of CUNFFT computation using a thread-level parallel mode in GPU. According to the scale of NDFT matrix, the HP-NFFT method may dynamically balance the relationship between performance and throughput capacity via adjusting the number of computer nodes used for parallel computations.

In the application of HP-NFFT method in calculating Madelung constant of a fluorite crystal structure and electrostatic interactions between charged ions in fluorite crystal structures with varied structure sizes, the HP-ENUF method is featured with high computational precision and excellent efficiency especially for super large simulation systems. From these computational results of benchmarking calculations, the HP-NFFT and the HP-ENUF methods exhibit strong competitiveness both in computational precision and in computational efficiency, and even more distinct for handling large NDFT matrices with at super extended scale.

**Acknowledgments**

This work is supported by the National Natural Science Foundation of China, and the Science and Technology Research Project Fund of Education Department of Jilin Province, China (Grant No. JJKH20190696KJ). Y.-L. Wang acknowledges the financial support from the Knut and Alice Wallenberg Foundation.